\documentclass[preprint]{aastex631}
\usepackage{amsmath}
\usepackage{booktabs}
\usepackage[shortlabels]{enumitem}
\usepackage{threeparttable}

\hypersetup{linkcolor=red,citecolor=blue,filecolor=cyan,urlcolor=magenta}

\begin{document}

\title{Se-ResNet+SVM Model: An Effective Method of Searching for Hot Subdwarfs from LAMOST.  
}

\author {Cheng Zhongding}
\affiliation{ School of Mechanical, Electrical \& Information Engineering, Shandong University, Weihai, 264209,  Shandong, China}

\author{Kong xiaoming}
\affiliation{ School of Mechanical, Electrical \& Information Engineering, Shandong University, Weihai, 264209,  Shandong, China}
 \email{xmkong@sdu.edu.cn}

\author{Wu Tianmin}
\affiliation{ School of Mathematics and Statistics, Shandong University, Weihai, 264209,   Shandong, China}

\author{Bu Yude}
\affiliation{ School of Mathematics and Statistics, Shandong University, Weihai, 264209,   Shandong, China}
\email{buyude@sdu.edu.cn}
\author{Lei Zhenxin}
\affiliation{Key Laboratory of Stars and Interstellar Medium, Xiangtan University, Xiangtan 411105,   China}
\affiliation{Physics Department, Xiangtan University, Xiangtan, 411105,  China}
\author{Zhang Yatao}
\affiliation{ School of Mechanical, Electrical \& Information Engineering, Shandong University, Weihai, 264209,  Shandong, China}

\author{Yi Zhenping}
\affiliation{ School of Mechanical, Electrical \& Information Engineering, Shandong University, Weihai, 264209,  Shandong, China}
\author{Liu Meng}
\affiliation{ School of Mechanical, Electrical \& Information Engineering, Shandong University, Weihai, 264209,  Shandong, China}



\begin{abstract}
In this paper, we apply the feature-integration idea to fuse the abstract features extracted by Se-ResNet with experience features into hybrid features and input the hybrid features to the Support Vector Machine (SVM) to classify hot subdwarfs. Based on this idea, we construct a Se-ResNet+SVM model, including a binary classification model and a four-class classification model. The four-class classification model can further screen the hot subdwarf candidates obtained by the binary classification model. The F1 values derived by the binary and the four-class classification model on the test set are 96.17\% and 95.64\%, respectively. Then, we use the binary classification model to classify 333,534 non-FGK type spectra in the low-resolution spectra of LAMOST DR8 and obtain a catalog of 3,266 hot subdwarf candidates, of which 1,223 are newly determined. Subsequently, the four-class classification model further filters the 3,266 candidates, 409 and 296 are newly determined respectively when the thresholds are set at 0.5 and 0.9. Through manual inspection, the true number of hot subdwarfs in the three newly-determined candidates are 176, 63, and 41, and the corresponding precision of the classification model in the three cases are 67.94\%, 84.88\%, and 87.60\%, respectively. Finally, we train a Se-ResNet regression model with MAE values of 1212.65 K for $T_{\rm{eff}}$, 0.32 dex for $\log$ g and 0.24 for [He/H], and predict the atmospheric parameters of these 176 hot subdwarf stars. This work provides a certain amount of samples to help for future studies of hot subdwarfs.

\end{abstract}


\section{Introduction} \label{sec:intro}
Hot subdwarf stars are a class of small-mass stars in the late evolutionary stages, which can be divided into two categories, B-type (sdBs) and O-type (sdOs), depending on their spectral types. In the H-R diagram, hot subdwarfs are located at the blue end of the horizontal branch extension and are therefore known as extreme horizontal branch stars. Although the mass of hot subdwarf stars is only half the mass of the Sun, they have an effective temperature of over 20,000 K and are considered to be an important source of ultraviolet radiation \citep{2009ARA&A..47..211H, 2002MNRAS.336..449H}. The peculiar characteristics of hot subdwarf spectra make the study of hot subdwarfs of crucial significance. It enables us to understand the stellar structure and evolution, globular clusters, and galaxy formation \citep{2009ARA&A..47..211H}. In addition, the prevalence of hot subdwarfs in the Milky Way helps to illustrate the UV-upturn phenomenon in elliptical galaxies \citep{1996A&A...313..405B}.

The study of hot subdwarf stars has been going on since the last century. For instance, \citet{1991PhDT........10S} proved the hypothesis that these stars are composed of burning helium cores, \citet{1997A&AS..125..501J} proposed a classification criterion for helium-rich subdwarfs, and \citet{2001ASPC..226..408S} suggested that sdB stars can be classified into three groups according to various radial velocities. 
But so far, the origin and evolutionary status of hot subdwarf stars remain a puzzle. Thus, it is essential to conduct a large sky survey to broaden the sample of hot subdwarfs for analysis.
With the release of the SDSS DR1 data, researchers have discovered more than three hundred hot subdwarf stars through spectroscopic observations and color discrimination, greatly expanding the sample of hot subdwarfs \citep{2003AJ....126.1023H, 2004ApJ...607..426K}. Other projects, such as the MUCHFUSS project \citep{2012ASPC..452..129G, 2013A&A...559A..35O}, the GALEX survey \citep{2012MNRAS.427.2180N}, and the Arizona-Montreal Spectroscopic Program \citep{2014ASPC..481...83F} also made contributions to the search for hot subdwarfs and the estimation of their stellar atmospheric parameters.
Using the SDSS colors and catalogs, a sample of 166 hot subdwarf stars was found from LAMOST DR1 through spectral analysis \citep{2016ApJ...818..202L}, indicating that LAMOST has great potential to carry out studies on the origin of such special stars as hot subdwarfs. 
Moreover, by combining Gaia large-scale survey data and LAMOST spectral data, color-magnitude maps with parallax and color information provided by Gaia can be used to select hot subdwarf candidates more efficiently from the huge amount of data \citep{2018ApJ...868...70L, 2019ApJ...881....7L, 2020ApJ...889..117L, 2022yCat..22560028L, 2022arXiv221112323L}. In addition, \citet{2021ApJS..256...28L} identified 1,578 hot subdwarf stars from the Gaia DR2 candidate catalog using LAMOST DR7 spectra and offered the atmospheric parameters of these stars. 
 \\ \indent
 The machine-learning method has been applied to search for the hot subdwarf stars from large spectra data. \citet{2017ApJS..233....2B} proposed that most of the hot subdwarfs were discovered by color cutting and visual inspection, which was inefficient for processing data with inhomogeneous photometric information derived from LAMOST. Consequently, the Hierachical Extreme Learning Machine (HELM) was introduced and they found approximately 10,000 hot subdwarf candidates from the LAMOST DR1 catalog.
\citet{2019ApJ...886..128B} presented a novel hybrid model to search for hot subdwarfs, combining the convolutional neural network (CNN) with a support vector machine (SVM). Experiments showed that it had advantages over other algorithms like LDA,  KNN, and HELM, and was proved to perform better in dealing with spectral data of LAMOST DR4. 
\citet{2022ApJS..259....5T} used a hybrid CNN-based approach, consisting of an eight-class classification method and a binary classification model, to identify hot subdwarf stars with an accuracy of 87.42\% and discovered 25 new hot subdwarf stars from LAMOST DR7-V1.  

Since the spectra of the hot subdwarfs are similar to those of main sequence B-type stars and BHB stars, it is crucial to develop efficient methods to discriminate hot subdwarfs from this kind of star. Hence, in this paper, we propose a hybrid method, combining Se-ResNet    (squeeze-and-excitation residual network) and SVM to extract spectral features and categorize hot subdwarfs. The central idea of the method is to integrate the abstract features extracted by Se-ResNet with experience features and then input the hybrid features to the SVM classifier. A binary classification model and a four-class classification model are constructed. Additionally, the Se-ResNet method is also used to estimate atmospheric parameters ($T_{\rm{eff}}$, $\log$ g, [He/H]) of the hot subdwarfs. The experimental results demonstrate that our model can efficiently identify hot subdwarfs, which provides a solid basis for searching for hot subdwarfs from large surveys in the future.

The rest of this article is organized as follows. The next section describes the dataset we use for the experiments and the methods for data pre-processing. Section \ref{sec:method} introduces the Se-ResNet and SVM model. 
Specific results of the binary and four-classification model can be seen in Section \ref{sec:exp}. Section \ref{sec:dis} presents the results of the search for hot subdwarfs in more than 300,000 low-resolution spectra and lists part of their parameters. A brief conclusion is provided in Section \ref{sec:con}.

\section{Data} \label{sec:data}
\subsection{Spectra} \label{subsec:spectra}
LAMOST, also known as Guo Shoujing Telescope, is located at Xinglong Observatory in the northeast of Beijing, China. LAMOST is a special reflecting Schmitt telescope with a wide field of view (diameter of $5^{\circ}$) and a huge effective aperture (4-6 meters, depending on the pointing height and hour angle). There are 4,000 optical fibers on the 20 square degrees focal plane. LAMOST not only provides us with a large number of stellar spectra but also allows multiple observations of the same object. It can collect 4,000 spectra simultaneously in a single exposure, making it a powerful spectroscopic survey telescope for a wide field of view and large-sample astronomy. It is planned to capture low-resolution ($R\sim1800$) optical spectra ($\lambda3,700-9,000\mathring{A}$) of more than 10 million stars in a continuous region of the sky ($-10^{\circ}$ to $90^{\circ}$ decl.) with a limiting magnitude of $R\sim18$mag \citep{2012RAA....12.1197C, 2012RAA....12..723Z}. So far, LAMOST DR8 Low-resolution Sky Survey has released more than 10 million low-resolution spectra in the wavelength range of 3,690$\mathring{A}$ to 9,100$\mathring{A}$.

\subsection{Labeled Data} \label{subsec:labdata}
Our task is to search for hot subdwarfs in the low-resolution spectra of LAMOST DR8 and estimate their atmospheric parameters ($T_{\rm{eff}}$, $\log$ g, [He/H]). In order to construct a classifier and prediction model, we need to first collect known hot subdwarf and non-hot subdwarf samples. Therefore, we obtain 5,837 spectra by cross-matching the spectra of LAMOST DR8 with 294 hot subdwarfs from \citet{2018ApJ...868...70L}, 388 hot subdwarfs from \citet{2019ApJ...881..135L}, 182 hot subdwarfs from \citet{2020ApJ...889..117L}, 5,874 hot subdwarfs from \citet{2020A&A...635A.193G}, 1,587 hot subdwarfs from \citet{2021ApJS..256...28L}, and 25 hot subdwarfs from \citet{2022ApJS..259....5T}. Then 2,422 spectra of hot subdwarf stars were obtained by removing the spectra with repeated OBSID and signal-to-noise ratio (SNR) less than 10, which are used as positive samples of the classifier. Among them, 1,774 hot subdwarf samples are provided with atmospheric parameters ($T_{\rm{eff}}$, $\log$ g, [He/H]) \citep{2020A&A...635A.193G}.

For the negative sample data set, we select seven types of celestial spectra, including 116 O-type stars, 2,000 BHB-type stars, 2,000 B-type stars, 2,000 A-type stars, 2,000 galaxies, 2,000 QSOs, and 2,000 UNKNOWN-type stars, 12,116 spectral data in total. Unlike the sample selection in \citet{2019ApJ...886..128B} and \citet{2022ApJS..259....5T}, considering the obvious characteristic differences between the spectra of hot subdwarfs ($T_{\rm{eff}}\geq$20,000 K) and FGK-type stars (3,900 K $\leq T_{\rm{eff}}\leq$ 7,500 K), and the high-precision parameters have been given by LAMOST for FGK-type stars, as well as a large number of FGK-type stars, we can remove the FGK-type stars labeled by LAMOST from the sample before conducting the search for hot subdwarfs. Therefore, the spectra of FGK-type stars are not taken into account in the construction of the negative sample data set.

To try to avoid mixing the selected A and B-type data sets with hot subdwarf samples, we randomly select 2,000 A-type and 2,000 B-type spectra from the LAMOST DR7 searched by \citet{2021ApJS..256...28L} and download in LAMOST DR8. The spectra of 2,000 BHB-type stars are randomly selected by cross-matching the BHB-type star catalogs provided by \citet{2021ApJ...912...32V} and \citet{2008ApJ...684.1143X} with the LAMOST DR8 catalog. As for the galaxy-type, UNKNOWN-type, and QSO-type spectra with rich samples, we randomly selected 2,000 of them in LAMOST DR8, respectively. However,  limited by the small number of  O-type spectra,  we only select 116 spectra in LAMOST DR8. The final positive and negative sample dataset is constructed with a ratio of about 1:5. Since the ratio of positive and negative samples is in a reasonable range and the proportion of O-type stars in the sample to be searched is small, we didn't perform data augmentation on the dataset. The dataset we constructed is shown in Table \ref{tab:table1}. 

The  SNR  of the spectra used for training and testing in the experiments is greater than 10, and the wavelength range of the spectra is 4,000$\mathring{A}$ to 8,000$\mathring{A}$, with a total of 3,909 data points.

\begin{table}[htbp]
  \centering
  \caption{The labeled data sets}
  \label{tab:table1}
  \setlength\headheight{10pt}
  \begin{tabular}{ccc}
\toprule
Data Sets   & Number of Samples    & Function    \\ \midrule
Positive Training Set & 1695 & Training classifiers \\
Positive Test Set  & 727   & Testing classifiers   \\
Negative Training Set   & 8481   & Training classifiers  \\
Negative Test Set   & 3635   & Testing classifiers  \\ 
\bottomrule 
  \end{tabular}
\end{table}

\subsection{Data Pre-processing} \label{subsec:preprocess}
The data can be pre-processed using the following steps:
\begin{enumerate}[(1)]
 \item Wavelength correction with radial velocity: the wavelength is corrected using the following equation,
\begin{equation}
w_{new}(n)=\frac{w(n)}{1+z},
\end{equation}
where $w(n)$ denotes the wavelength of the observed spectra, $z$ is the redshift of the observed spectra.

 \item Min-Max normalization: The fluxes of each spectrum are normalized to the range [0,1] according to the following equation,
\begin{equation}
	x^\prime=\frac{x-x_{min}}{x_{max}-x_{min}},
\end{equation}
where $x$ denotes the flux of the observed spectra, $x_{min}$ and $x_{max}$ denotes the minimum and maximum flux of the spectra.
\end{enumerate}

\section{Method} \label{sec:method}
Well-chosen input features can improve classification accuracy substantially, or equivalently, reduce the amount of training data needed to obtain the desired level of performance \citep{forman2003extensive}.

When selecting the input data for the model, we can manually choose several important feature bands of hot subdwarfs as input, based on prior knowledge. It is also possible to choose to input the entire spectrum and entirely rely on machine learning algorithms to extract features. The abstract features learned by deep learning methods have a good generalization capability and also allow us to obtain implicitly deeper features. Inspired by the work of \citet{SHI2020105219} and \citet{9064540}, this paper applies a hybrid feature method to combine the experience features with the abstract features extracted by machine learning algorithms, which considering both the effectiveness of the experience features and the generalization ability of the abstract features.

Se-ResNet \citep{hu2018squeeze} is the winning model of the 2017 ImageNet competition, and experiments show that Se-ResNet has a stronger learning ability than traditional CNN networks and ResNet. Se-ResNet not only retains its own residual blocks but also enables the model to learn the importance of different features by importing a squeeze excitation module to focus more on those more significant features. Therefore, Se-ResNet is chosen as the neural network model to extract abstract features in this study.

SVM, first proposed by \citet{cortes1995support} in 1995, is a widely used binary classification model. It maps the linearly indistinguishable or hard-to-distinguish sample data points in low-dimensional space to high-dimensional space via the Kernel trick, making them linearly distinguishable. SVM is used in this paper for classification and identification due to its good performance in the classification of small sample data.

 In this paper, we combine Se-ResNet and SVM, using Se-ResNet to extract the abstract features of the spectra, then fusing the abstract features with the experience features into hybrid features, and finally using SVM for classification. To begin with, we briefly introduce Se-ResNet and SVM, followed by a detailed description of the architecture of the model.
\subsection{Se-ResNet} \label{subsec:se}
\begin{figure}[htbp]
		\centering
	\includegraphics[width=0.9\linewidth]{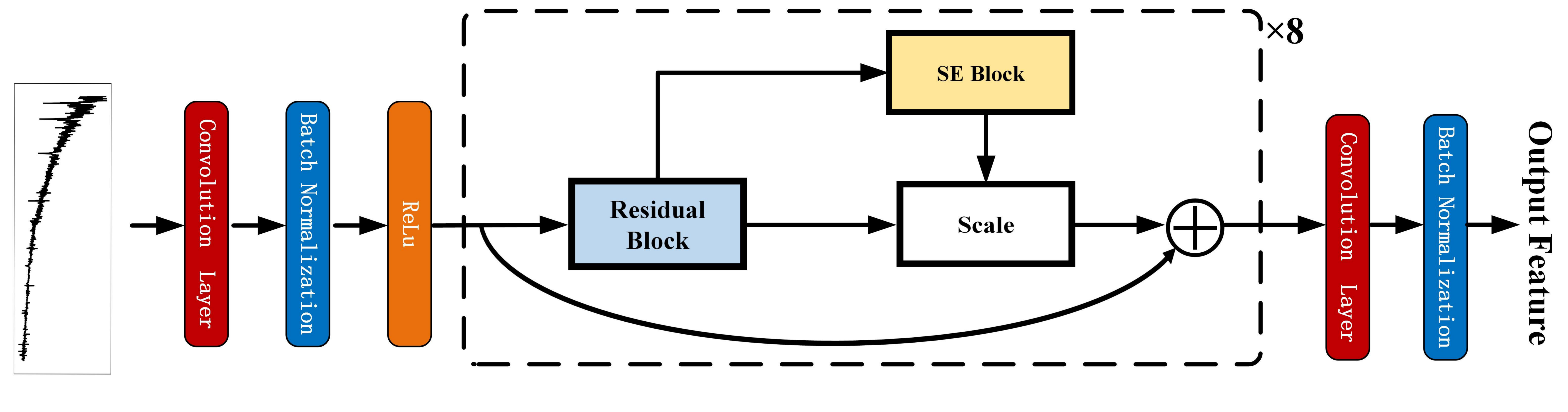}
		\caption{The structure of the Se-ResNet used in this paper.}
       \label{fig:model}
\end{figure}

The 1-D Se-ResNet network used in this paper is composed of one convolutional layer, one batch-normalization layer, one ReLu layer, and eight residual blocks. Each residual block is followed by a squeeze and excitation network (SE-Block). Figure \ref{fig:model} displays the architecture of the network. Each residual block consists of two convolutional layers, two batch-normalization layers, and one ReLu layer; the Se Block includes two consecutive fully-connected layers, one global pooling layer, and one sigmoid activation function layer. The structure of the residual block and Se Block is shown in Figure \ref{fig:block} below. To obtain the abstract features, a 1$\times$512 feature matrix is exported after the final batch-normalization layer to derive the implied abstract features of this inputting spectrum.

\begin{figure}[htbp]
		\centering
	\includegraphics[width=0.9\linewidth]{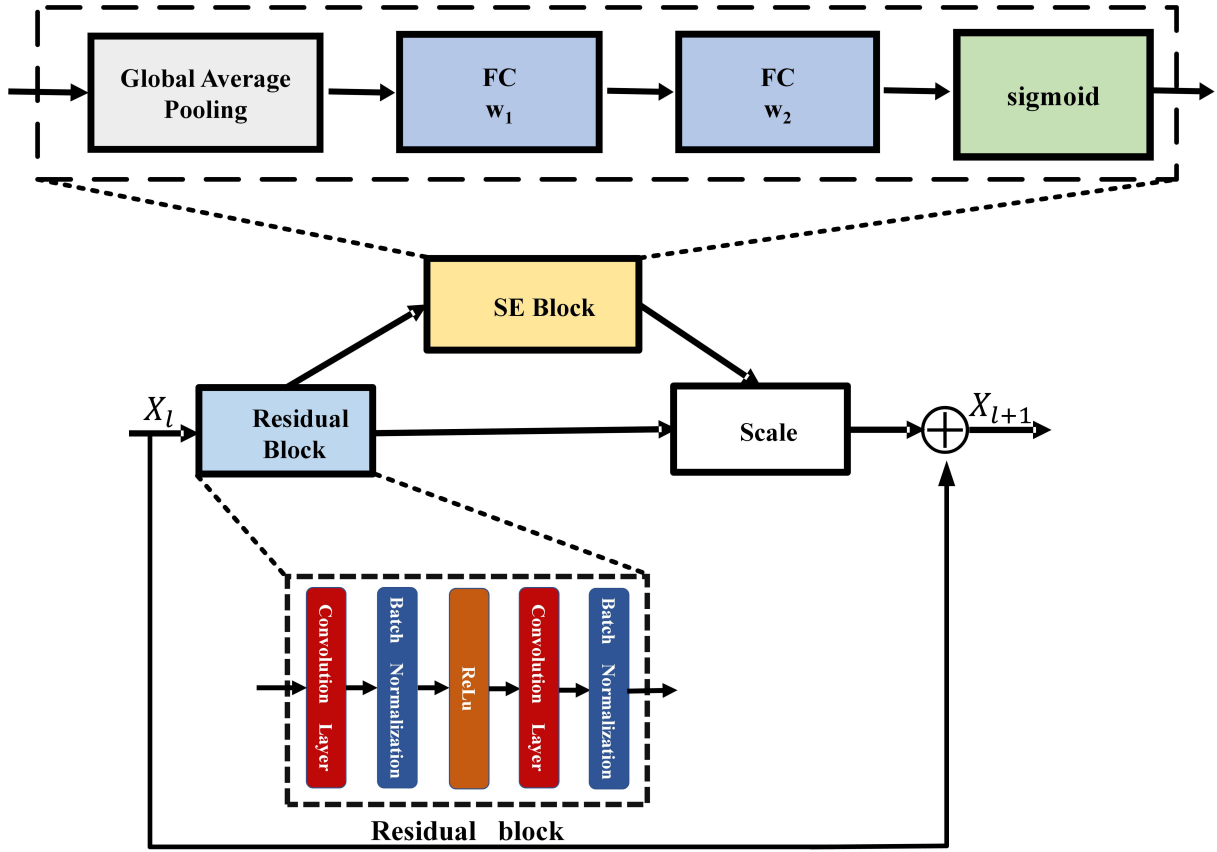}
		\caption{The structure of the residual block and Se Block.}
       \label{fig:block}
\end{figure}

The residual block is responsible for implementing the shortcut connection to build a direct mapping between different layers. The role of the shortcut connection is to ensure that the value of the gradient is at least one when it is propagating, which can eliminate the phenomenon of gradient disappearance. This process satisfies the following equation:
\begin{equation}
X_{l+1}=ReLu(F_{i}(X_{l})+F_{j}(X_{l})),
\end{equation}
where $X_{l+1}$ represents the output features, $X_{l}$  is the input features, the function $F_{i}$ is the direct mapping of $X_{l}$, $F_{j}$ denotes the indirect mapping of $X_{l}$, and ReLu is responsible for mapping the input neurons to the output. The squeeze excitation module can learn the importance of different feature channels. The traditional convolutional process considers the weight of each feature channel to be identical, but in fact, the importance of different feature channels is different. Adding a Se Block after the residual block can make the network pay more attention to the feature channels with high weight and less attention to the unimportant feature channels. The squeeze and excitation network computes the weight of each feature channel in the following procedure.

Firstly, Squeeze performs a global pooling of each feature channel of the 1-D spectral data according to the following formula to obtain $N$ average values:
\begin{equation}
Z_{c}=F_{sq}(u_{c})=\frac{1}{H\times W}\sum_{i=1}^{H}\sum_{j=1}^{W}u_{c}(i,j),
\end{equation}
where $Z$ is the weight, the subscript $c$ denotes the feature map, $F_{sq}$ denotes the procedure function used to derive the weight mentioned above, $H$, $W$ denote the height and width of the feature map, respectively, and $u_{c}(i,j)$ is the value of the $i$-th row and $j$-th column of the feature map.

Secondly, Excitation regards the value after the Squeeze operation as the initial attention weight and inputs it into the two fully-connected layers, generating the sigmoid output. Excitation operates as the following equation:
\begin{equation}
s=F_{ex}(Z,W)=\sigma(W_{2}ReLu(W_{1}Z)),
\end{equation}
where $Z$ is the values obtained by Squeeze, $W_{i}$ is the fully-connected layer, $\sigma$ is the sigmoid function.

The role of Scale is to multiply the weight generated by Se Block with the original feature matrix to obtain a new feature matrix. The new feature matrix has the following characteristics. When the weight of a featured channel becomes larger after passing the Se Block, then the value of the new feature channel after this multiplication will increase and the impact on the final output will become larger.  Accordingly, when the weight becomes smaller, then the value of the feature channel will become smaller and the impact on the final output will also become smaller. 

SE-ResNet is trained using back-propagation and optimized using Adam algorithm \citep{kingma2014adam}. The learning rate is set to 0.01. Since the work in this paper involves a binary classification problem, the cross-entropy function is used as the loss function.

After feeding the 1-D stellar spectra into the network, the 1$\times$512 abstract feature is extracted using the 1-D Se-ResNet. Representing the abstract feature by $x$, the abstract feature vector is $[x_{1},x_{2},x_{3},\dots,x_{512}]$.Table 2. shows the structure and parameters of the abstract feature extraction model

\begin{table}[htbp]
  \centering
  \caption{Network architecture and output parameters.}
  \label{tab:table8}
  \setlength{\tabcolsep}{3mm}
  \begin{tabular}{cccc}
\toprule
NO  & Layer   & Output size    \\ \midrule
0   & Input   & 1$\times$ 3909  \\
1   & Convolution-1   & 64$\times$3909  \\
2   & BatchNormalization-1   & 64 $\times$3909  \\
3   & ReLu   & 64 $\times$ 3909  \\
4   & Se-ResNet Block-1   & 64$\times$3909  \\
5   & Se-ResNet Block-2   & 64$\times$ 3909  \\
6   & Se-ResNet Block-3   & 128$\times$1955  \\
7   & Se-ResNet Block-4   & 128$\times$1955  \\
8   & Se-ResNet Block-5   & 256$\times$ 978  \\
9   & Se-ResNet Block-6   & 256$\times$ 978  \\
10   & Se-ResNet Block-7   & 512$\times$489  \\
11   & Se-ResNet Block-8   & 512$\times$489 \\
12   & Convolution-2   & 1$\times$512  \\
13   & BatchNormalization-2   & 1$\times$512  \\
\bottomrule 
  \end{tabular}
\end{table}

\subsection{Feature Fusion} \label{subsec:fusion}
\begin{figure}[htbp]
		\centering
	\includegraphics[width=0.9\linewidth]{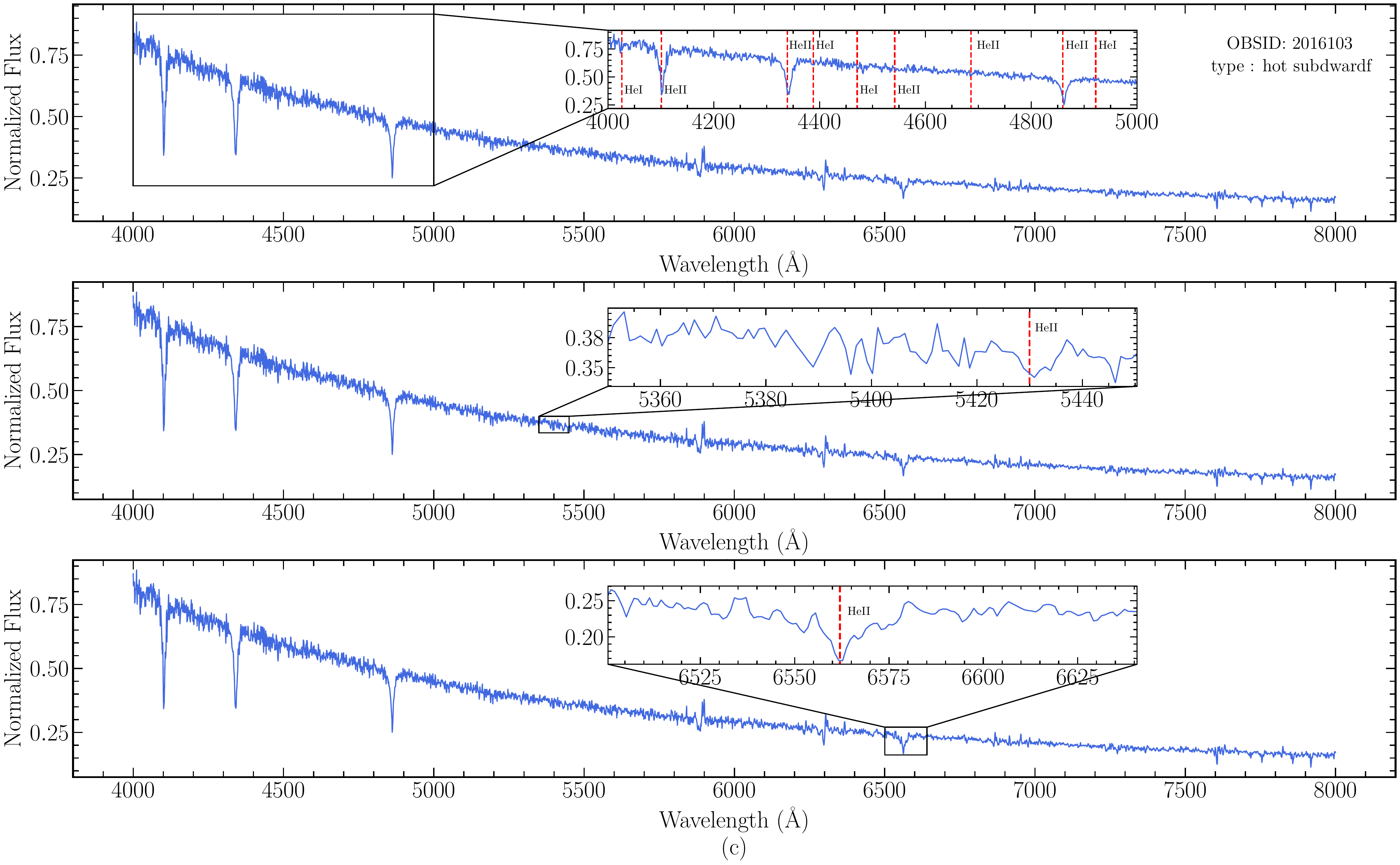}
		\caption{The example spectrum, the red dotted line is the absorption line at the corresponding position of HeI and HeII.}
       \label{fig:spectrum}
\end{figure}

According to \citet{2021ApJS..256...28L} and \citet{2018ApJ...868...70L}, three wavelength ranges are chosen as the experience features, 4,000$\mathring{A}$ to 5,000$\mathring{A}$, 5,350$\mathring{A}$ to 5,450$\mathring{A}$, and 6,500$\mathring{A}$ to 6,640$\mathring{A}$. As shown in Figure \ref{fig:spectrum}, these three feature bands encompass all the important characteristic absorption lines of hot subdwarfs. A total of 1,213 feature points are sampled in these three wavelength ranges.

Using $y$ to represent the abstract features, the experience feature vector can be represented as $[y_{1},y_{2},y_{3},\dots,y_{1213}]$.

Afterward, fusing the abstract features and experience features, we can obtain the hybrid feature vector of length 1,725, denoted as $[y_{1},y_{2},y_{3},\dots,y_{1213},x_{1},x_{2},x_{3},\dots,x_{512}]^T$. Assuming that the input of the classifier is $N$ and the number of the spectra is $y^1,\dots,y^N$, the following matrix is the input to the classifier in this experiment:
\begin{equation}
\begin{pmatrix}
y_{11}^1 & \cdots &\cdots &y_{11}^N \\
\vdots & & &\vdots\\
y_{1213}^1&\ddots  & &y_{1213}^N\\
x_{1}^1&  & \ddots&x_{1}^N\\
\vdots & & &\vdots\\
x_{512}^1&\cdots  &\cdots &x_{512}^N
\end{pmatrix}
\end{equation}

\begin{figure}[htbp]
		\centering
	\includegraphics[width=0.9\linewidth]{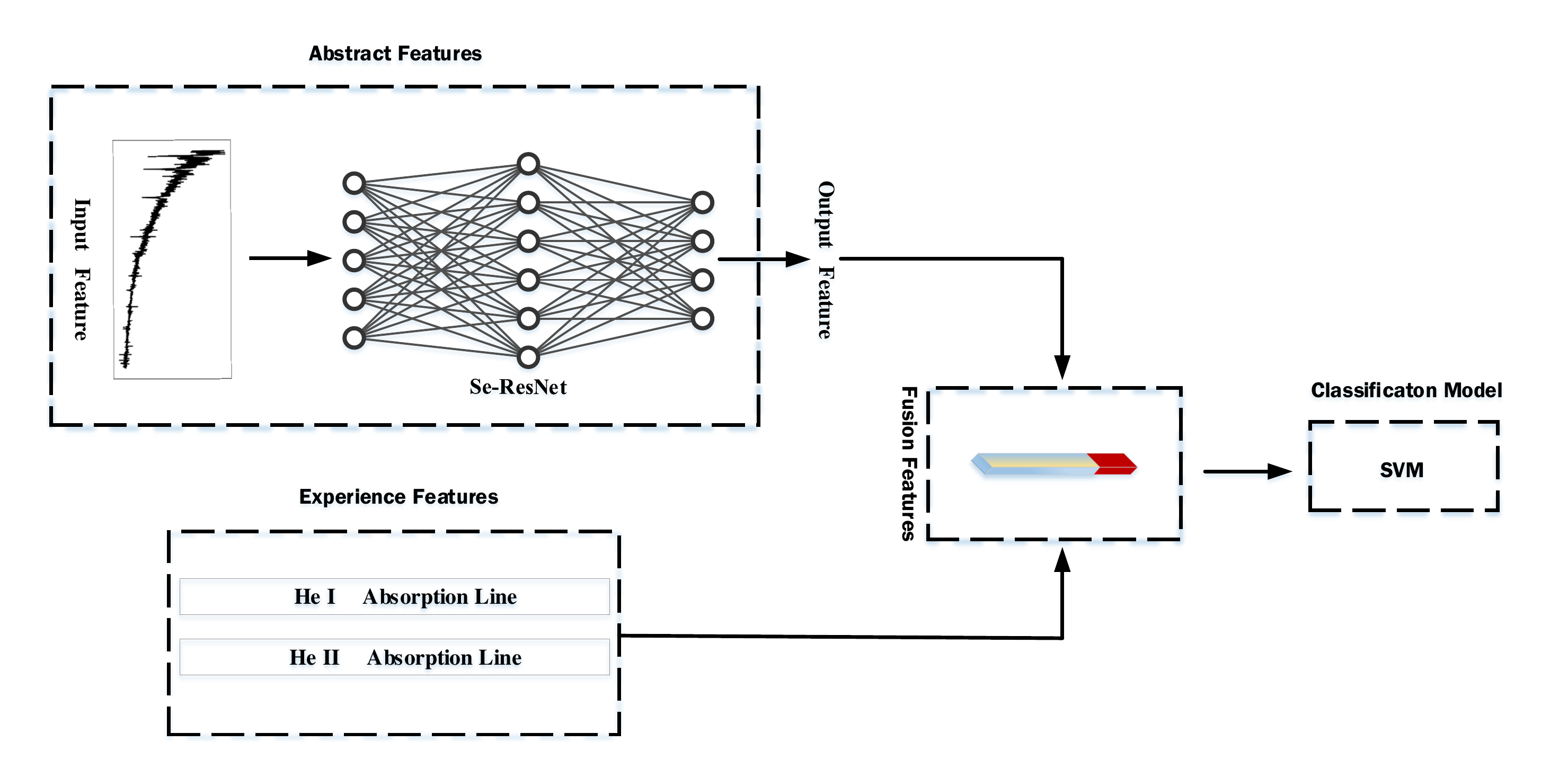}
		\caption{The flow chart of the method.}
       \label{fig:flow}
\end{figure}

Finally, we use the SVM classifier to train and classify, deriving the classification results. The flow chart of the method is shown in Figure \ref{fig:flow}.

\subsection{SVM} \label{subsec:svm}
SVM is a classification model that has been widely used in astronomical research \citep{2018InPhT..95...61C,2008AIPC.1082..144H,2012ApJ...760...15F,peng2012selecting,2014NewA...28...35B,2015MNRAS.453..122S}. Since the RBF Gaussian radial basis kernel function has maintained a stable performance in most studies and can fit data better for nonlinear mapping \citep{2017InPhT..86...23Y,pattanayak2021novel}, the RBF kernel function is adopted in the work.

The SVM model has two very important parameters $C$ and $\gamma$, where $C$ is the penalty factor, i.e., the tolerance for error. The higher the $C$, the more intolerant the error is and the more likely it is to be overfitted. The smaller the $C$, the smaller the $C$, and the more likely the model will be under-fitted. $\gamma$ determines the distribution of the data mapped to the new feature space. The larger the $\gamma$, the fewer the support vectors, and the smaller the $\gamma$, the more the support vectors. The number of support vectors affects the speed of training and prediction. We use the grid search method to find the optimal values of $\gamma$ and $C$ and find that the model is optimal when $\gamma$ equals 0.001 and $C=10,000$.

\section{Experiment} \label{sec:exp}
\subsection{Performance metric}
We apply three indicators ($Precision$, $Recall$, $F1\_score$) to assess the performance of the classification model. The definitions are as follows:
\begin{eqnarray}
Precision=\frac{TP}{TP+FP},\\
Recall=\frac{TP}{TP+FN},\\
F1\_score=\frac{2\times Precision\times Recall}{Precision+Recall},
\end{eqnarray}
where $TP$ denotes the number of hot subdwarfs correctly predicted as hot subdwarfs by the classifier, $FN$ denotes the number of hot subdwarfs incorrectly predicted as non-hot subdwarfs, $TN$ is the number of non-hot subdwarfs correctly classified as non-hot subdwarfs, and $FP$ is the number of non-hot subdwarfs incorrectly classified as hot subdwarfs.
$Precision$ is the probability that the samples predicted to be hot subdwarfs are true hot subdwarfs. $Recall$ is the probability of being predicted as hot subdwarfs in the hot subdwarf samples, and $F1\_score$ is the harmonic mean of $Precision$ and $Recall$.

In the model of estimating the parameters of stars, we evaluate the performance of the model using the mean absolute error (MAE), and the formula is as follows:
\begin{equation}
MAE=\frac{1}{n}\sum_{i=1}^{n}(|y_{i}-\hat y_{i}|),
\end{equation}
where $n$ is the number of samples in test set, $y_{i}$ is the true value of the $i$-th data, $\hat y_{i}$ is the predicted value of the $i$-th data.

\subsection{Classification Construction}


In this section, we constructed a two-stage classification model including a binary classification model and a four-class classification model to classify hot subdwarfs. In the binary classification stage, 2,422 pre-processed positive sample spectra and 12,116 negative sample spectra introduced in section \ref{subsec:labdata} were randomly divided into a training set and a test set in the ratio of 7:3. The training set is used to train the Se-ResNet model to continuously tune all parameters of the neural network mentioned in Table \ref{tab:table8}, and the test set is used to evaluate and validate the performance of the model after training. The results on the test set are shown in Table \ref{tab:table2}, with a precision of 97.32\%, a recall of 95.04\%, and an F1\_score of 96.17\%.

\begin{table}[htbp]
  \centering
  \caption{The performance of the binary classification model.}
  \label{tab:table2}
  \setlength{\tabcolsep}{4mm}
  \begin{tabular}{cccc}
\toprule
Method  & F1\_score(\%)   & Precision(\%)  & Recall(\%)    \\ \midrule
Se-ResNet+SVM & 96.17 & 97.32  &95.04 \\
\bottomrule 
  \end{tabular}
\end{table}

Although the F1\_score of discriminating hot subdwarfs of the binary classification model can reach 96.17\% on the test set, in practical applications, faced with mass and complex real data to be identified, it is difficult for machine learning models trained on limited samples, especially small samples, to obtain the results that are comparable to the test set, resulting in a higher than expected contamination rate for new candidate samples. 
Hence, a four-class classification model was constructed to further filter the hot subdwarf samples found by the binary classification model.

\begin{figure}[htbp]
	\centering
	\includegraphics[width=0.9\linewidth]{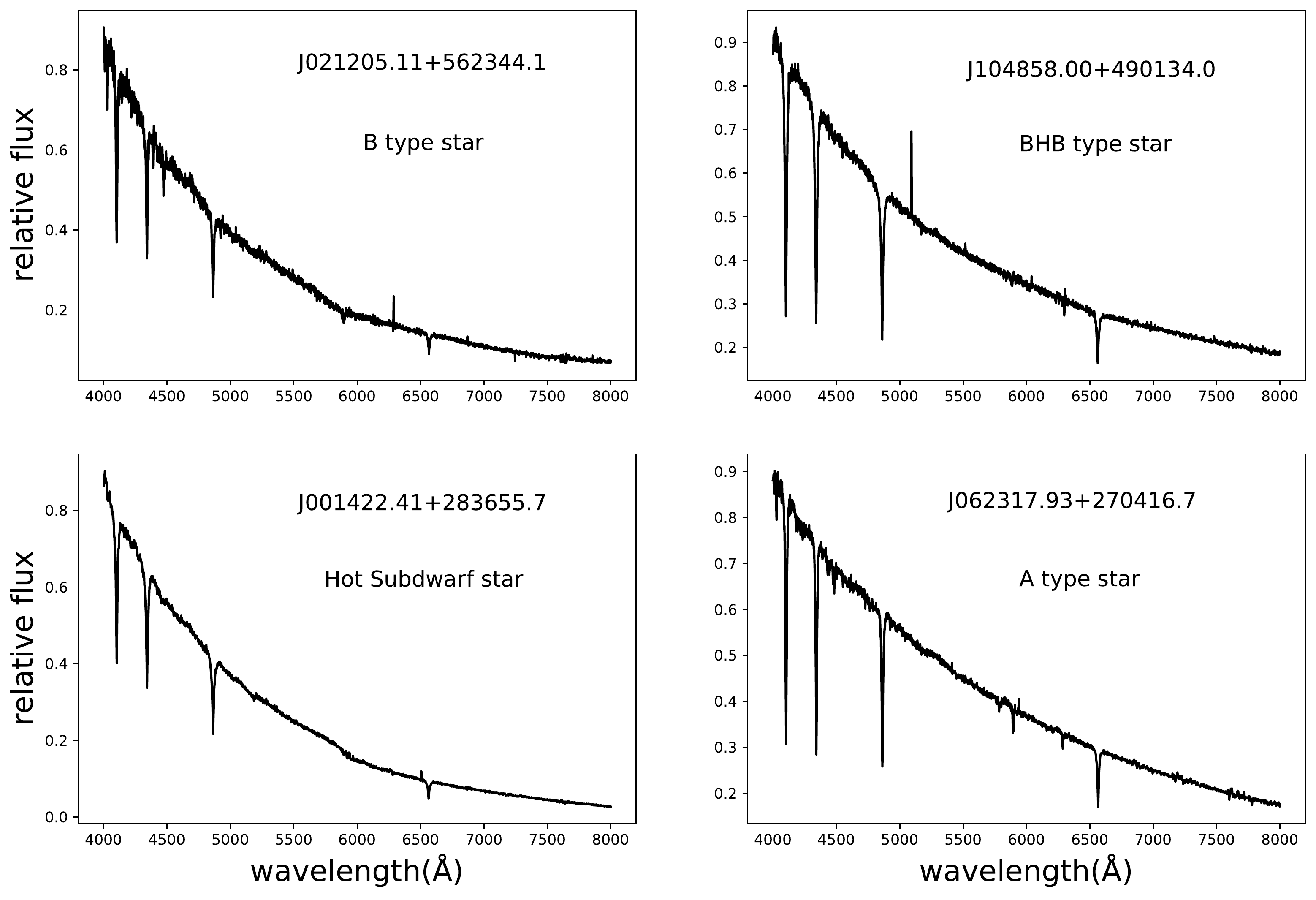}
		\caption{Example spectra selected from LAMOST data. Top left: example spectrum of a B-type star. Top right: example spectrum of a BHB-type star. Bottom left: example spectrum of a hot subdwarf star. Bottom right: example spectrum of an A-type star.}
    \label{fig:spec}
\end{figure}

For the negative samples of the four-class classification model, we selected A-type, B-type, and BHB-type spectra that have similar characteristics to the hot subdwarfs and thus are more likely to be misclassified. 2,000 spectra were randomly re-selected for each type following the same selection criteria as the binary classifier samples, for a total of 6,000 negative samples. Figure \ref{fig:spec} shows the spectra of these four types of stars. After data pre-processing, a four-class classification model was constructed using the same algorithmic steps.

After the same process described above for abstract feature extraction and feature reconstruction, the data was finally fed to the SVM for multi-classification training. The F1\_score is 95.64\%, the precision is 96.13\%, and the recall is 94.72\%, as shown in Table \ref{tab:table3}.

\begin{table}[htbp]
  \centering
  \caption{The performance of the four-class classification model.}
  \label{tab:table3}
  \setlength{\tabcolsep}{4mm}
  \begin{tabular}{cccc}
\toprule
Method  & F1\_score(\%)   & Precision(\%)  & Recall(\%)    \\ \midrule
Se-ResNet+SVM & 95.64 & 96.13  &94.72 \\
\bottomrule 
  \end{tabular}
\end{table}

\subsection{Stellar parameter estimation  } \label{subsec:pc}

Due to the high temperature and few absorption lines of hot subdwarfs, estimating the stellar parameters of hot subdwarfs is difficult. Preliminary experiments have shown that it is hard to obtain satisfying estimating results by constructing a model for hot subdwarfs with a small sample (1,774 data) of labeled data. Hence, for the estimation model, we added a data preprocessing step, which proved to have a key impact on the performance of the prediction algorithm. We adopt a polynomial fitting method based on the spline function for normalization to eliminate the influence of pseudocontinuum on the predictor. The details of this procedure are as follows.
\\Step 1: Remove and discard the local minimal and discrete points.  \\
Step 2: Normalize the spectral data by the maximum value, i.e. $flux = flux /\max(flux)$.
\\ Step 3: Fit each spectrum using the spline function to obtain the fitted spectrum. 
\\Step 4: Divide the observed spectrum by the fitted spectrum to obtain the normalized spectrum. 

\begin{figure}[htbp]
	\centering
	\includegraphics[width=0.9\linewidth]{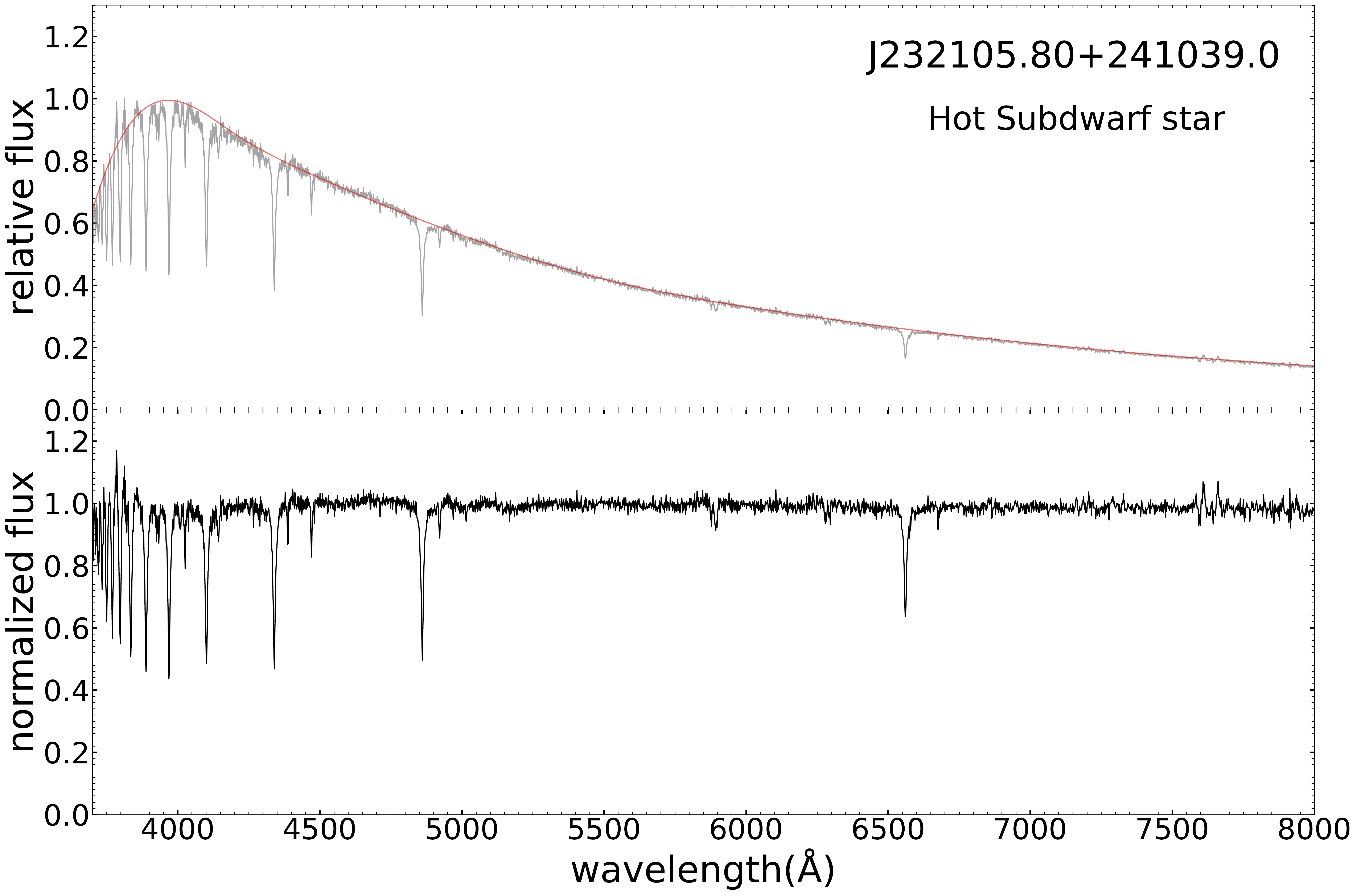}
		\caption{LAMOST observation spectrum (spec-55878-B87802\_2\_sp01-068) and its normalization spectrum. The original LAMOST spectrum is shown above, and the continuous normalized spectrum is shown below.}
       \label{fig:norm}
\end{figure}

Figure \ref{fig:norm} illustrates the normalized spectra obtained using the method. Additionally, we select the wavelength range from 4,000$\mathring{A}$ to 6,000$\mathring{A}$ as the input features.

\begin{figure}[htbp]
	\centering
	\includegraphics[width=0.9\linewidth]{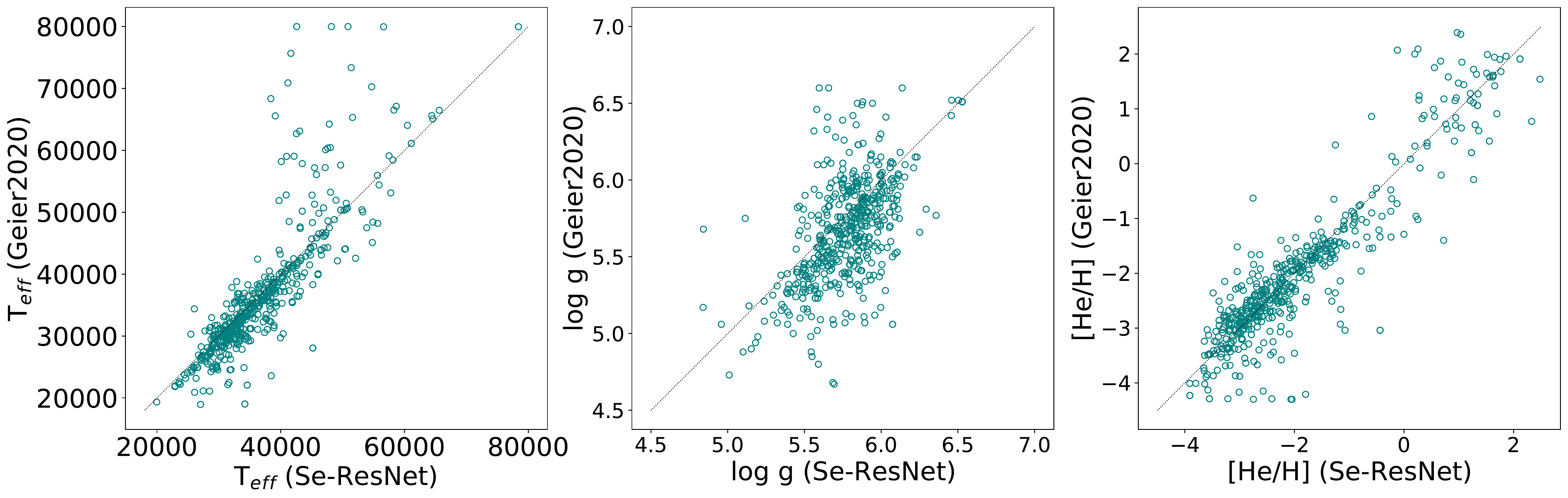}
		\caption{Comparison of the stellar parameters predicted by Se-ResNet in this work with target values calculated by Geier2020. The dotted lines above are one-to-one lines.}
       \label{fig:prediction}
\end{figure}

In this paper, the Se-ResNet algorithm is used to estimate the atmospheric parameters of hot subdwarfs. The difference with the structure of Se-ResNet for extracting abstract features is that instead of outputting directly after the batch-normalization layer, we add a ReLu layer and a global pooling layer as well as a fully-connected layer for outputting the results.

Similarly, we randomly divide the data into training and test sets in the ratio of 7:3 and input them to the SE-ResNet for training. After the loss function is stabilized, the MAE values of $T_{\rm{eff}}$, [He/H], and $\log$ g of the hot subdwarfs are 1212.65 K, 0.32 dex, and 0.24 dex, respectively. The distribution of stellar parameters of hot subdwarfs predicted by the Se-ResNet algorithm with those determined by \cite{2020A&A...635A.193G} is present in Figure \ref{fig:prediction}.

\section{Results and Discussion} \label{sec:dis}

\subsection{Comparison with Other Methods}

Before using the Se-ResNet+SVM algorithm to retrieve hot subdwarfs, we investigated the effect of different input features of the spectra on the performance of the algorithm. In addition to the hybrid-feature input method, we also adopted the full-spectrum input (4,000$\mathring{A}$-8,000$\mathring{A}$) and feature-band input (4,000$\mathring{A}$-5,000$\mathring{A}$, 5,350 $\mathring{A}$-5,450$\mathring{A}$, 6,500$\mathring{A}$-6,640$\mathring{A}$) methods separately for comparison. For these two feature input methods, the Se-ResNet algorithm was used to build a classifier instead of an abstract feature extractor, so the SVM classifier is no longer needed. The difference between the Se-ResNet networks used in the comparison experiments is that after the batch normalization layer, we did not output abstract features but added a fully connected layer for classification. As shown in Table \ref{tab:table4}, we can see that the classification effect has been significantly improved after the application of the  Se-ResNet+SVM algorithm.

\begin{table}[htbp]
  \centering
  \caption{The comparison results of the binary classification model using different input features and methods.}
  \label{tab:table4}
  \setlength{\tabcolsep}{4mm}
  \begin{tabular}{cccc}
\toprule
Method  & F1\_score(\%)   & Precision(\%)  & Recall(\%)    \\ \midrule
Se-ResNet(feature band) & 93.47 & 93.54  &93.41 \\
Se-ResNet(full-spectrum) & 89.87 & 93.60  &86.42 \\
Se-ResNet+SVM(hybrid feature) & 96.17 & 97.32  &95.04 \\
\bottomrule 
  \end{tabular}
\end{table}

In the parameter prediction phase described in Section 4.3, after applying the normalization method based on the spline function on the input features, the performance of the parameter prediction modal has improved significantly. Therefore, we also apply this normalization method to the classification modal to observe its impact on the classifier performance. Considering that in the first classification stage, the binary classification searches for hot subdwarfs in massive spectra, it is time-consuming to normalize these massive data. However, in the second classification stage, the four classifiers only need to further filter the small initial candidate samples obtained in the first stage, so, we only compared the effect of using the normalization method based on the spline function and the Min-Max normalization method on the four-class classification model. Table \ref{tab:table5} shows that there are negligible differences in the classification outcomes. Thus, in the stage of identifying hot subdwarfs, we still adopt the optimal normalization method for the classification.

\begin{table}[htbp]
  \centering
  \caption{The performance of the four-class classification model using different normalization methods.}
  \label{tab:table5}
  \setlength{\tabcolsep}{4mm}
  \begin{tabular}{cccc}
\toprule
Method  & F1\_score(\%)   & Precision(\%)  & Recall(\%)    \\ \midrule
Normalization method based on spline function & 95.70 & 96.37  &95.04 \\
Min-Max normalization method & 95.64 & 96.13 &94.72 \\
\bottomrule 
  \end{tabular}
\end{table}

\subsection{Searching for Hot subdwarfs from LAMOST DR8}
In the LAMOST DR8 database, we download a total of 333,534 low-resolution spectra based on the filtering conditions of SNRU$\geq$5 and non-FGK-type stars, and after the same data pre-processing steps, we apply the binary classification model to the spectra and obtain 3,266 hot subdwarf candidate spectra. Among them,   2,043 spectra are the hot subdwarf samples in the training set and 1,223 are the newly-determined hot subdwarf candidates by the model. Subsequently, we apply the four-class classification model to these 3, 266 spectra, and the results show that 2,289 spectra are identified to be hot subdwarfs candidates with a threshold of 0.5.  Among them, 1,880 are hot subdwarf samples in the training set. After manual verification, 63 of the other 409 spectra are newly discovered hot subdwarfs, and 346 are other types of spectra that are misclassified as hot subdwarfs. The precision of identifying hot subdwarfs is 84.88\% (precision=(1,880+63)/2,289) using the two-step classification model. 

With a threshold of 0.9, 2,058 of the 3,266 spectra are classified as hot subdwarfs, of which 1,803 are actual hot subdwarfs (41 are newly discovered hot subdwarf samples and 1,762 are hot subdwarf samples in the training set), and 255 are other types of spectra that are misclassified as hot subdwarfs. The precision of identifying hot subdwarfs is 87.61\%(precision=(1,762+41)/2,058).

We also manually verify the 1,223 hot subdwarf candidates obtained by the binary classifier and find that there are 176 new hot subdwarfs. The precision of identifying hot subdwarfs using the binary classifier alone is 67.94\%(precision=(2,043+176)/3,266). 

As is shown above, the accuracy of classifying hot subdwarfs is significantly improved in the dataset filtered by the four-class classification model. However, simultaneously, the recall rate also decreases apparently.

Applying the parameter estimation model in Section \ref{subsec:pc}, we measure the atmospheric parameters ($T_{\rm{eff}}$, $\log$ g, and [He/H]) for the 176 newly discovered hot subdwarfs. Table \ref{tab:table6} lists all parameters of part of the 176 newly discovered hot subdwarfs, including obsid, designation, S/N, R.A., dec, $T_{\rm{eff}}$, $\log$ g, and He abundance.

\begin{table}[htbp]
  \centering
  \caption{The Atmospheric Parameters of 176 newly identified hot subdwarfs in this work.}
  \label{tab:table6}
  \setlength{\tabcolsep}{3mm}
  \begin{tabular}{cccccccc}
\toprule
obsid  & designation   & SNRu  & Ra(degree)& Dec(degree)&$T_{\rm{eff}}$(K)&$\log$ g(dex)&[He/H](dex)\\ \midrule
644815154	&J144546.39+234845.9	&111.28	&221.44	&23.81	&32224.27	&5.76	&-2.68\\
545404103	&J061340.67+225236.8 &46.21&	93.42&	22.88&	27499.99&	5.49	&-2.69\\
208816099	&J052142.56+082530.7	&14.88	&80.43	&8.43 	&31168.91	&5.78	&-2.71\\
269511194	&J000324.47+584628.0	&35.49	&0.85 	&58.77	&28106.37  &5.56&-2.71\\
282604062	&J001920.59+463306.8 &68.78     &4.84 	&46.55    &25572.93   &5.71&-1.74\\
\bottomrule 
  \end{tabular}
\end{table}
 
\section{Conclusion} \label{sec:con}
In this study, we construct a hybrid Se-ResNet+SVM feature model for classifying hot subdwarfs from LAMOST DR8 and build a prediction model for atmospheric parameters. The data-balanced approach for data augmentation of sparse numbers of stars is not used in this work. Applying the Se-ResNet+SVM to LAMOST DR8 spectra, we obtained the results indicating that Se-ResNet+SVM has better performance on the classification of hot subdwarfs in small sample datasets. This study provides an alternative way to search for hot subdwarfs in large sky surveys. Moreover, we address the problem of difficulty in distinguishing B-type, A-type, and BHB-type stars from hot subdwarfs in model training, and create a four-class classification model to better discover hot subdwarfs. Subsequently, we use the model to search for hot subdwarfs from the LAMOST DR8 database, and most of the hot subdwarfs can be correctly identified. The experiments reveal that although 87.60\% of the hot subdwarfs predicted by the final four-class classification model prove to be hot subdwarfs, the four-class classification model still greatly improves the accuracy of finding hot subdwarfs compared to the binary classification model. However, in practical applications, it will lead to the loss of some samples that are actually hot subdwarfs. These 176 newly discovered hot subdwarfs will provide a reference for subsequent studies in related fields. 
The atmospheric parameters of 176 newly identified hot subdwarfs in this work are available at https://github.com/chengLLLLLLLLLL/new-hot-subdwarf-stars-identified-from-lamost-DR8.

~\\
\indent
This work is supported by the National Natural Science Foundation of China (NSFC) under Grant Nos.11873037, U1931209, and 11803016, the science research grants from the China Manned Space Project with No. CMS-CSST-2021-B05 and CMS-CSST-2021-A08, 
and is partially supported by the Young Scholars Program of Shandong University, Weihai (2016WHWLJH09). Lei acknowledges support from the Natural Science Foundation of China No. 12073020, the Scientific Research Fund of Hunan Provincial Education Department grant No. 20K124, Cultivation Project for LAMOST Scientific Payoff and Research Achievement of CAMS-CAS, the science research grants from the China Manned Space Project with No. CMS-CSST-2021-B05.

\bibliography{ref}{}
\bibliographystyle{aasjournal}



\end{document}